\documentstyle[12pt]{article}
\newcommand{\beq}{\begin{equation}}
\newcommand{\eeq}{\end{equation}}

\def\md{\hbox{$m_D$ }}

\def\gtorder{\buildrel > \over {_{\sim}}}

\begin{document}
\hoffset = -1truecm
\voffset = -2truecm

\null
\vskip 5mm

\rightline{IC/93/359}
\rightline{October 1993}

\vskip 5mm

\begin{center}
{\large \bf  NEUTRINO MASSES AND MIXING}
\vskip 1cm

{ALEXEI YU. SMIRNOV}\footnote{on leave from Institute for
Nuclear Research, Russian Academy of Sciences, Moscow, Russia, \\
\phantom{reku}e-mail: smirnov@ictp.trieste.it}\\
{\it International Centre for Theoretical Physics, Strada Costiera 11,}

{\it 34100 Trieste, Italy}\\

\vskip 1cm

{ABSTRACT}
\end{center}
\begin{list}{}
{\setlength{\leftmargin}{12mm}
\setlength{\rightmargin}{12mm}
\setlength{\baselineskip}{4mm}
\footnotesize
\item}
Restrictions on the neutrino masses and lepton mixing  are reviewed. Solar,
atmospheric and relic neutrinos give the indications of  existence of
nonzero neutrino masses and mixing. The data
pick up two regions of mixing angles which are or appreciably larger
or appreciably smaller than the Cabibbo angle. Some theoretical schemes
with {\it large} or {\it small} lepton mixing are discussed.
\end{list}

\vskip 1cm

\noindent
{\large \bf 1. Negative Results}

\vskip 2mm
Let us mark the frontiers of present knowledge.

{\it Electron (anti)neutrino}: three tritium experiments,
Los Alamos$^1$, Mainz$^2$, and Livermore$^3$  have overcome 10 eV barrier of
the  upper bounds. The strongest limits
are at the level
7 - 8 ~ eV:
\beq
m(\bar{\nu}_e) <  \left\{\matrix {7.2~ {\rm eV}~~ 95\%~ C.L. & {\rm Mainz}^2
\cr
                                  8  ~ {\rm eV}~~~ 95\%~ C.L. & {\rm
Livermore}^3 \cr } \right. .
\eeq
Both experiments give small (about
$1 \sigma$ below 0) negative values of $m^2$.
This may testify for some undiscovered systematic error, which could
slightly relax the above limits. Zurich$^4$ limit is 11 eV (95\% C.L.),
and  $m^2$ perfectly
agrees  with zero.

Double beta decay searches restrict the effective
Majorana mass of the  electron neutrino; the best limit has been
obtained
by Heidelberg-Moscow collaboration$^5$ with enriched isotope $^{76}$Ge:
\beq
m_{ee} \equiv \sum_{i} \eta^{CP}_{i} |U_{ei}|^2 m_i  < 1.3 ~~ {\rm eV},
{}~~ 90\% C.L. .
\eeq
Geochemical experiment with $^{128}Te / ^{130}Te$ gives$^6$ similar
restriction:  $m_{ee} < (1.1 - 1.5)$ eV. In (2)
$U_{ei}$, $m_i$, and $\eta^{CP}_i$ are the admixture
in the electron neutrino state, the mass and the CP- parity
of the i-component of neutrino. If all components have the  same
CP-parities the limit from double beta decay strengthens the
bound  (1),  in case of opposite parities the cancellation
in (2) allows even for lightest component to have a mass   of the order
of 7 eV.\\

{\it Muon neutrino}: a restriction on the mass follows from
measurements
of the momentum of the muon in the pion decay  $\pi^+
\rightarrow \mu^+ \nu_{\mu}$ ($\pi^+$ at the rest)
and from independent measurements of  the pion mass$^7$.
Method of the determination of
$m_{\pi}$
has an ambiguity:  two different values of $m_{\pi}$ where obtained.
One  value
results in  strongly
negative (about $5\sigma$ below zero)
 $m^2(\nu_{\mu})$,
whereas another
one  gives
$m^2 (\nu_{\mu}) = - 0.002 \pm 0.030~ {\rm MeV}^2$ in agreement with zero.
This corresponds to  a new upper bound$^7$
\beq
m(\nu_{\mu}) < 220 ~{\rm kev} ~~~ 90 \% ~C.L.
\eeq

{\it Tau neutrino}: the invariant masses of five pions in
the  decay $\tau \rightarrow 5 \pi + \nu_{\tau}$ are measured
and an upper limit on the tau neutrino mass follows from spectrum of these
invariant masses
near the  end point:
\beq
m(\nu_{\tau}) <  \left\{ \matrix{ 31~ {\rm MeV} ~~ 95 \%~ C.L. & {\rm ARGUS}^8
\cr
                             32.6~ {\rm MeV} ~~ 95 \%~ C.L. & {\rm CLEO}^9
\cr  } \right. .
\eeq
CLEO has detected 60 events with 5 charged pions and 53 events
with three charged and two neutral pions. The limit
is determined by  few  events near the end point, and
ARGUS group is lucky: its limit was obtained with just 20 (5 charged
pions) events  (the probability of this is$^9$ $p \approx 0.04$).

Strong limits on tau neutrino mass follow from
a primordial nucleosynthesis$^{10}$.
At the epoch of
nucleosynthesis,
the neutrinos with
masses $ > 1$ MeV were nonrelativistic.
Their contribution to the energy density
in the Universe is the interplay of two factors: the
mass and the concentration, $n(T)$;
the latter is exponentially suppressed at $T < m$. As the result
of the interplay the energy density, $m \cdot n(T)$,
has a maximum at $m$ = (4
- 6) MeV
which is equivalent to the  contribution to energy density of the $\Delta
N_{eff} = (4 - 5)$
massless neutrinos. The effective number of the relativistic neutrinos,
$N_{eff}$, is
restricted by present $^4$He abundance.
The limit $N_{eff} < 3.4$ excludes$^{10}$  the region of  masses
(0.5 - 25) MeV  for the Majorana neutrinos
with a lifetime larger than 1 s; for $\tau > 10^3~ s$ the
excluded region is (0.5 - 32) MeV. Recently the bounds have been refined.
In$^{11}$  using even  weaker restriction
$N_{eff} < 3.6$
the region 0.5 - 35 MeV was excluded for $\tau > 10^2$ s.
The limit $N_{eff} < 3.3$ results$^{12}$ in the the exclusion interval
0.1 - 40 MeV for
$\tau_{\nu} > 10^3~ s$
. This means that
for stable
neutrinos  there is no gap
between laboratory (4) and NS bounds and upper limit is pushed down to
\beq
m(\nu_{\tau}) < 0.1 - 0.2~ {\rm MeV}~~~~ (\tau > 10^3 s)
\eeq
Fast decay
$\nu_{\tau} \rightarrow \nu ' + J$, where $J$ is the majoron, can
relax the bounds. It was found$^{12}$ that the gap appears for
$\tau_{\nu} < 10^{2}~ s$
. At $\tau_{\nu} < 10^{-2} s$ the mass region  3 - 30 MeV is not
excluded.
The gap for stable neutrinos appears if
one admits $N_{eff} > 4$.

The above result has a number of implications.
A relatively  stable tau neutrino is appreciably lighter than electron.
There is no decay
$\nu_{\tau} \rightarrow e^+  e^- \nu '$. If the
mass of $\nu_{\tau}$ is generated by the see-saw mechanism then the
corresponding mass of the right handed component should be of the
order of $m_{\tau}^2/ m(\nu_{\tau}) \approx 3 \cdot 10^4$ GeV,
which appreciably larger than the electroweak scale, so that the
electroweak see-saw does not work etc.

The above limit from the nucleosynthesis is applied also to the muon
neutrino.\\

Mixing of the neutrinos  has a number of consequences:
kinks on the Kurie plot of beta decays;
additional peaks in energy distributions of charged leptons from two
body decays, e.g.  $\pi \rightarrow \mu \nu_{\mu}$,
neutrino decays, oscillation of neutrinos.
Up to now no peaks, no kinks, no decays, no oscillations$^{13}$ have been
found,
which  results in  upper bounds on mixing as function of neutrino mass/
mass squared difference .
The most dramatic event here is second comming
(with 0.7 - 1.2 $\%$ admixture) and second death of
the 17 kev neutrino. Some recent restrictions (at 95$\%$ C.L.)  on the
admixture  of the heavy  neutrino in the electron neutrino state
are summarized below
$$
|U_{eh}|^2 < \left\{
\begin{array}{lll}
0.073 \% &   17~ {\rm kev}         & INS ~{\rm Tokyo}^{14} \cr
0.15 \%  &   10.5 - 25~ {\rm kev}  & --''--    \cr
0.11 \%  &   17 ~  {\rm kev}       & {\rm Zurich}^{15} \cr
0.25 \%  &   10 - 45~ {\rm kev}    & {\rm Argonne}^{16} \cr
0.29 \%  &   17~ {\rm kev}         & {\rm Princeton}^{17}\cr
0.24 \%  &   17 ~    {\rm kev}     & {\rm Oklahoma}^{18} (99\%) \cr
\end{array}
\right .
$$
Let us remark that kev region is still interesting: models
developed in the context of the 17 kev neutrino predict
naturally much smaller mixing than it was found in the "positive"
experiments. One may keep in mind $|U_{eh}|^2 \sim m_e/m_{\tau}
\approx 0.03 \%$ or $(m_{\mu}/m_{\tau})^2 \approx 0.25 \%$ etc. \\

In conclusion,
 the gap (in log scale)  between the
upper bounds on neutrino masses  and charged leptons is much
larger than a
spread  between masses  of charged fermions from the same generation:
$\frac {m(\nu_l)_{upper}}{m_l} \ll \frac {m_l}{m_q}$.
The mixing of the neutrinos with masses near the upper bounds is
strongly restricted.

\vskip 5mm
\noindent
{\large \bf 2. Positive Results
Mixing}

\vskip 3mm

At present three observations testify for
nonzero neutrino masses  and mixing.\\

{\it The solar neutrinos}: all the experiments
(Homestake, Kamiokande, SAGE, \\GALLEX)
show the deficit of
$\nu_e$-flux$^{19, 20}$. Moreover, there  is an  indication on energy
dependence
of the suppression which cannot be explained by any astrophysical reasons.
(After GALLEX publications it becomes fashionable to discuss the
``detection" solution of the problem, keeping in mind that some of the
experimental results may have  wrong interpretation).

The data obtained so far can be perfectly described by the
resonant flavor conversion (MSW-effect)$^{21}$ $\nu_e \rightarrow
\nu_{\mu} (\nu_{\tau})$ with small vacuum mixing (see fig. 1):
\beq
\Delta m^2 = (0.5 -1.2) \cdot 10^{-5}~ {\rm eV}^2,~~~
\sin^2 2\theta = (0.3 - 1.0) \cdot 10^{-2}
\eeq
or large vacuum mixing:
\beq
\Delta m^2 = (1 - 3) \cdot 10^{-5}~ {\rm eV}^2,~~~
\sin^2 2\theta = (0.65 - 0.85)
\eeq
for two neutrino case$^{22}$. In presense of  third
neutrino the  allowed domains become larger, in particular,  the
region of  small mixing solutions can be extended$^{23}$ up to
$\sin^2 2\theta = 8 \cdot 10^{-4}$ and
$\Delta m^2 = 8 \cdot 10^{-5}~ {\rm eV}^2$.

Alternatively the data can be described by long length vacuum oscillations
(``just-so")$^{24}$ with parameters$^{25}$:
\beq
\Delta m^2 = (0.7 - 1.0) \cdot 10^{-10}~ {\rm eV}^2,
{}~~~
\sin^2 2\theta = 0.85 - 1.0,
\eeq
but this solution  is disfavored
by SN1987A data (see below).\\

{\it Atmospheric neutrinos}:
the  {\it double ratio}
($\frac{\mu-like}{e-like})_{observation}$/($\frac{\mu-like}{e-like})_
{MonteCarlo}$
for the contained events  measured by
water \`{C}erenkov detectors Kamiokande and IMB  is $^{26}$ about
0.6, i.e. appreciably smaller than expected 1.
The results from the
iron calorimeters (FREJUS, NUSEX) give for the double ratio a value which
agrees  with 1 (although the errors are rather large).
First results
from  new iron calorimeter experiment SOUDAN-II are$^{27}$ not yet
decisive:
the double ratio, being about 0.7, agrees with that of the Kamiokande, but it
is just 1.5 $\sigma$ below  1. These results can be explained by the
oscillations
$\nu_{\mu} - \nu_e$ or
$\nu_{\mu} - \nu_{\tau}$ or
$\nu_{\mu} - \nu_s$, where $\nu_{s}$ is the sterile neutrino,
although the last possibility is disfavored by the nucleosynthesis
consideration.
The ratio of the {\it stopping to through-going muons} from down semisphere
(IMB) is in agreement with predictions without any oscillations. The fluxes of
the {\it upward going
muons} measured by Kamiokande, IMB and Baksan are also in  agreement with
predictions (but the theoretical uncertainties are estimated to be
of the order of $40\%$). Negative results give the exclusion region of
the neutrino parameters which however does not cover all the region of the
positive results. The survival domain is
\beq
\Delta m^2 = (0.3 - 3) \cdot 10^{-2}~ {\rm eV}^2,~~~
\sin^2 2\theta = 0.4 - 0.6
\eeq
for $\nu_{\mu} - \nu_{\tau}$ oscillations (fig.1) and
\beq
\Delta m^2 = (0.3 - 2) \cdot 10^{-2}~ {\rm eV}^2,~~~
\sin^2 2\theta = 0.35 - 0.8
\eeq
for
$\nu_{\mu} - \nu_e$
.  In the indicated regions the
data from different experiments are described at  about $2\sigma$
level and, consequently, the total probability that all the
data are fitted by the parameters (9, 10) is rather small.
Let us stress that maximal mixing is excluded as a solution of the
atmospheric
neutrino problem by Frejus result on the double ratio and
by IMB result on stopping/through going muons, and the uncertainties of
both results are rather small.
This fact is very important for theoretical implications.\\

{\it Relic neutrinos}: a formation of  large scale structure of the Universe
implies the  existence of hot dark matter component which contributes
up to
30$\%$ of the critical density. The neutrino with mass
\beq
m \sim 2 - 7~~ {\rm eV}
\eeq
can play the role of such a component$^{28}$ . \\

In conclusion,  let us stress that  mixing angles corresponding
to  positive results (6 - 10)
do
not coincide with
Cabibbo-Kobayashi-Maskawa mixing angles. In particular, mixing of
$\nu_{\mu}$ and  $\nu_e$ should be or appreciably {\it smaller} or
appreciably {\it larger} than the Cabibbo mixing. In next
sections we will discuss the possible inference of this result.

\vskip 0.8cm
\noindent
{\large \bf 3. Small Mixing}
\vskip 3mm

Let us start with one observation.
The angles in the region of small mixing solution of the solar neutrino
problem (6) are
a little bit smaller than
the root squared from the ratio of the electron ($m_e$) and the muon
($m_{\mu}$) masses:
\beq
\theta_{e \mu} < \theta_l \equiv  \sqrt {\frac{m_e}{m_{\mu}}} .
\eeq
($\sin^2 2\theta_l \sim 0.02$).
One can correct the above expression by adding some contribution from the
neutrino mass ratio:
\beq
\theta_{e \mu} = \left |\sqrt \frac{m_e}{m_{\mu}} - e^{i\phi}
\theta_{\nu} \right |,
\eeq
where $\phi$ is a phase. Such a relation between the angles and the masses
is similar to the relation in  quark sector$^{29, 30}$ and
follows naturally from  the Fritzsch ansatz for mass matrices$^{30}$.
According
to this ansatz in a certain basis (which can be fixed by some family
symmetry) the mass matrices have the following form:
$$
\pmatrix{
0 & \mu & 0 \cr
\mu & 0 & m \cr
0 & m & M \cr
},
$$
and moreover $\mu \ll m \ll M$. \\

Smallness of neutrino masses can be reconciled with
quark-lepton symmetry as well as  with
possible similarity of
 mixing in  quark and lepton sectors  (13)
in framework of the see-saw mechanism of neutrino mass generation$^{31}$.
The see-saw  mechanism
 naturally relates a smallness of the neutrino
masses with  neutrality of neutrinos. According to this
mechanism
the Majorana mass matrix for left
(active) components ($\approx \nu_L$) equals
\beq
m^{maj} = - m_D M^{-1} m_{D}^{T}\ ,
\eeq
where \md is the Dirac mass matrix  and
$M$ is the Majorana mass matrix  of
right-handed neutrino components. It is supposed that
$M \gg m_{D} $.

An  attractive feature of this mechanism is that
the Dirac matrix, \md, can be similar (in scale and structure)
to that in the quark sector which is naturally implied by
 quark-lepton symmetry and Grand Unification. The essential difference
in scales of neutrino masses, and probably in lepton mixing,
follows from structure of expression (14) and from the
Majorana mass matrix.\\

According to (14),
to predict the neutrino masses and mixing
one should fix \md and M. Two remarks are in order.

1. Quark-lepton symmetry or/and Grand Unification
allow to relate the quark and the lepton mass matrices.
But simple equality $m_d = m_l$ is not true. Masses of quarks and
leptons from the same generation are different. Relation between these
masses  are well described by Georgy-Jarlskog ansatz$^{32}$ at GU scale:
$
m^0_d = 3 m^0_e, ~~~
m^0_s = \frac {1}{3} m^0_{\mu},~~~
m^0_b = m^0_{\tau}.
$~~~~
One should keep in mind also possible deviation from
the equality $m^0_D = m^0_{up}$. Large t-quark mass also
indicates that simple relation may not be true.

2. There is no analogy of $M$ matrix in  quark sector. Matrix
$M$ can be fixed or restricted if the right handed neutrino  components,
$\nu_R$,  enter the same multiplet as known quarks and leptons
(e.g. 16-plet of SO(10) ) and if $\nu_R$ as well as quarks and leptons
acquire  the masses from the interaction with the same Higgs multiplet
(e.g., 126 plet of SO(10)).
Mixing is the result of the {\it difference} in mass matrices of
up and down particles from the weak doublets.
The difference can not arise from the interaction with only one Higgs
multiplet. Therefore one should introduce at least two
multiplets , or suggest some additional sources of the fermion masses.

The structure of  $M$ could be restricted if this matrix is generated by
radiative
corrections$^{33}$. Also some special ansatz can be invented to fix $M$.

Thus, the  recipe of predicting the neutrino masses and mixing is the
following.

a). Fix the relation between the neutrino Dirac mass matrix, $m_D$, and
mass matrix of  quarks
 by choosing some GUT with certain (minimal) set of Higgs multiplets.

b). Use  the parameters of quark
mass matrices as well as the masses of charge leptons as an input, to
determine  the Dirac neutrino mass matrix.

c). Certain ansatz (e.g., Fritzsch ansatz) for a structure of mass
matrices can be suggested to diminishing a number of input parameters and
to make the predictions for the quark sector too.

d). Fix the matrix $M$ by some additional ansatz or
relate its structure with other  mass matrices by GUT.
(Overall scale of mass in M could be restricted by the unification scale
or by the intermediate scale of  symmetry violation).

d). Using formula (14) find the mixing and neutrino masses. \\

Let us review some results obtained along with this line.

The relation (13) can be reproduced  in the see-saw mechanism.
Suppose $m_l$ and $m_D$ have the Fritzsch
structure. Let $M$ be proportional to the unit matrix:
$M = M_0 \cdot I$ (fig.2, FR), then according to (14) the Majorana neutrino
mass matrix is diagonalized by the same transformation as diagonalizes
the neutrino Dirac matrix $m_D$. Consequently,  in the relation
(13),  in a direct analogy with quark sector
one gets the neutrino contribution
$\theta_{\nu} \sim \sqrt {m_{1D} / m_{2D}}$,
where $m_{1D}$ and $m_{2D}$
are the eigenvalues of the neutrino
Dirac mass matrix,
. Since
$M$ is proportional to the unit matrix, the real masses of light
neutrinos will have the quadratic hierarchy:
$m_i \sim (m_{iD})^2 / M_0$. The ratio of the Dirac masses can be
rewritten as $m_{1D} / m_{2D} =
\sqrt {m_{1} / m_{2}}$, and
for the neutrino contribution to the mixing angle one finds
\beq
\theta_{\nu} \sim ^4\sqrt { \frac{m_1}{m_2}}.
\eeq
For second and third generations one finds similarly
$\theta_{\nu} \sim ^4\sqrt { \frac{m_2}{m_3}}$.
So, even for relatively strong hierarchy of masses the
mixing
angle  can be rather large$^{34}$. This relation was used to reconcile
large $\mu - \tau$ mixing implied by the atmospheric neutrino deficit
and small mixing solution of the solar neutrino problem.
Indeed, for $m_2 = 3 \cdot 10^{-3}$ eV  and $m_3 = 10^{-1}$ eV one
gets  $\sin^2 2\theta = 0.5$ - precisely in the
desired region (9).

In fact, the above result
implies  a weak hierarchy of
eigenvalues
of  the  neutrino Dirac matrix: $m_{3D}/ m_{2D} < 6$. This realizes  a
problem for Gran Unification since
for up-quarks one has $m_t/m_c > 10^2$. \\

The above  analysis allows just to {\it relate} the angles and the
masses  but it does not allow to predict the masses.
The predictions can be made if one  assumes  certain {\it textures}
of mass matrices
(by puting some elements of the matrices to be zero)
{\it and}  also some  GUT relations.
Usually the SO(10) model is used with all the fermions from one family in one
16-plet.
The textures of the mass matrices in some suggested models are shown in
fig. 2 b-d. Matrices are supposed to be symmetric and the elements of
matrices are generated by  the interactions with 10- and 126- plets of
Higgs bosons. Different
10- and 126- plets have
different couplings in family
space and acquire
vacuum expectations in different directions. This (and also the
fact that  10-plets give $m_q = m_l$, whereas
126-plets generate $3 m_d = m_l$) allows to get a desirable
relations between nonzero elements of matrices.

Let us comment on some features of different models.

In$^{35}$ the Dirac mass matrices  have  the Fritzsch form
(see fig.2, BS).
Neutrino Dirac matrix  coincides with
matrix of up quarks at the unification scale
(both are generated by 10-plet).
Majorana matrix $M$ is related to the Dirac matrices $m_{d,l}$:
the same 126-plet gives the contribution to 23-elements of these
matrices.
An additional 126-plet is introduced to generate the
11-element of $M$ and thus to get nonzero
determinant of $M$.
(Both 10-plet
and 126-plet contribute to 23 elements of the charged leptons and
down quarks mass matrices to reproduce Georgi-Jarlskog ansatz).
The $e \mu$ element of the
neutrino mixing matrix equals
\footnote {Here and in further relations the masses of quarks and leptons
are at the GUT scale.}
\beq
\theta_{e \mu} \sim V_{2 e} = \sqrt {\frac{m_e}{m_{\mu}}} - e^{i\phi}
\sqrt {\frac{m_u}{m_c}}.
\eeq
The model predicts also observable $\nu_{\mu} - \nu_{\tau}$ oscillations
with $\sin^2 2\theta \sim 0.1$ and the mass of third neutrino (1 - 3) eV in
the cosmologically interesting domain.

 In $^{36}$ (fig.2, DHR)
the neutrino Dirac matrix and   $m_{u}$
are generated by 126-plets, so that
$m_D \propto m_u$. Structure of
$M$ is partly related to structure of $m_{u, D}$: $M$ has
minimal number of nonzero elements (12, 21, 33) and these elements are
generated by the same 126-plet that generates the corresponding elements of
$m_{u, D}$.
Mixing between second and third families is induced by third
126-plet which generate 23-elements of $M_{u, \nu}$.
(Fritzsch structure is modified and there
is another realization of
Georgi-Jarlskog
ansatz:  126-plet contribute to 22-element of $M_{e,d}$).
The $e \mu$ neutrino mixing  is found to be
\beq
\theta_{e \mu} = \left | \frac {m_e}{m_{\mu}} +
\frac{1}{9} \frac{m_u}{m_c} -
\frac{2}{3} \sqrt {\frac{m_e m_u}{m_{\mu} m_c}} \cos \phi \right
|^{\frac{1}{2}},
\eeq
where $m_1$ and $m_2$ are the majorana masses of  light neutrinos.
At $m_2 = 1.7 \cdot 10 ^{-3}$ eV the model predicts
$\sin^2 2\theta _{e \mu} = (1.7 \pm 0.2) \cdot 10 ^{-2}$ which
is a little bit larger than what is needed for the solution of the
solar neutrino problem. The tau neutrino mass is near to
the cosmologically interesting domain $\sim 0.4 - 0.6$ eV.
There is no  explanation of the atmospheric neutrino
deficit in this as well as in previous models . \\

In model$^{37}$ (fig.2, BS')
the matrices $m_D$ and $m_u$ as well as $m_{d,l}$
are generated both by 10-
and by 126- plets in such a way that mixing between second and third
generation is enhanced in lepton sector.  Matrix $M$ is partly
related to $m_{d, l}$: in both matrices 22 and 23 elements are generated by
the  same 126-plets.
As in$^{35}$ an additional
126-plet is introduced to generate 11-element of matrix $M$ and to
have nonzero determinant of this matrix.
(The element 22 of $M_{e,d}$ generated by 126-plet allows to
reproduce the Georgi-Jarlskog ansatz).
Mixing between two first lepton generations is determined by
\beq
\theta_{e \mu} \sim V_{2 e} = \cos \theta \left | \sqrt{\frac
{m_e}{m_{\mu}}} -
\frac {1}{9} e^{-i\phi} \sqrt {\frac {m_u}{m_c}} \right |.
\eeq
%Factor 1/9 is related to generation of 23 elements by 126-plets
%and zero value of 22-element  in $m_D$ matrix.
The parameter $\cos \theta$ determines
the mixing between the second and the third generations:
$
V_{3 \mu} = \left | 3 \epsilon \cos \theta + \sin \theta
e^{i\sigma } \right |
$,
where $\epsilon \equiv \frac{1}{4} \sqrt {1 - m_b^2 / m_{\tau}^2}$.
 The model allow to explain
both the solar and the atmospheric neutrino deficits.

Equations (15 - 18) give an idea on  the dependence of the neutrino
contribution  to mixing, $\theta_{\nu}$, on structure of mass matrices.\\

In the considered schemes the predictions of lepton masses/mixing
follow from proposed textures of the mass matrices and certain GUT
relations, but to get desirable relations
between nonzero elements one should introduce  several 10- and 126-
plets.   Another possibility to get the predictions is to use
only one 10-plet and only one 126-plet but do not assume  any
textures of mass matrices$^{38}$.  Doublet components of 126-plet
receive the induced vacuum expectations at tree level and thus
126-plet both generates the $M$ matrix and gives the
contributions to the Dirac mass matrices $m_{d,l}$ and $m_{u, D}$.
Several different sets of the parameters have been found depending,
in particular, on the ratio of the induced VEV's. In one case, e.g.,
$m_2 \sim 2\cdot 10 ^{-3}$ eV and mixing between the first and
the second generations is fixed by $\sin^2 2 \theta \approx
0.013$ allowing to solve the solar neutrino problem. Third mass
can be about 2 eV and its admixture in the light states is on the level
of the existing experimental bounds from the oscillation experiments.
Changing the  scale of the neutrino masses one can explain the
solar neutrino problem by $\nu_e \rightarrow \nu_{\tau}$ conversion.

Another approach to generate the fermion
 masses and to make the model more economical
have been proposed$^{39}$.
Desired structure of the mass matrices
can be reproduced by
the nonrenormalizable interactions
of fermion multiplet with small number of low dimension Higgs
multiplets.
Only 10- plet and the adjoint representation 45-plets
are used$^{39}$. Minimal set of the nonrenormalizable
operators has been chosen which results in the mass matrices similar
to those  shown
in fig.2 (BS'). The nonrenormalizable interaction operators allow to
explain not only the texture of the matrices but also the hierarchy of
their elements.

\vskip 0.8cm

\noindent
{\large \bf 4. Large Mixing}
\vskip 3mm

Large lepton mixing can be obtained in the
see-saw mechanism as a consequence of certain structure of the
matrix $M$. Two sets of the conditions have been found$^{40}$.

First implies strong mass
hierarchy in the heavy Majorana sector  and
certain correlation between the  Dirac and the Majorana mass matrices
of neutrinos, e.g.,
\beq
M \sim {1 \over \mu}  \cdot m_{D}^{T}  m_{D} .
\eeq
Here $\mu$ is some mass parameter.
Such a possibility has been marked in$^{41}$.
The above relation can be obtained if the right-handed neutrinos
acquire the Majorana masses  via the see-saw
mechanism too. This implies the existence of new heavy
neutral leptons  $N_L = (N_{1L}$, $N_{2L}$, $N_{3L})$, singlets of
the electroweak symmetry, with
bare Majorana masses, $m_{N}$, at some large scale
. The  Dirac mass terms of N and  $\nu_R$ can be  generated
by   the Yukawa interactions with
scalar singlet , $\sigma$, which  acquires
a vacuum expectation value $\sigma_0 \gtorder 10^{12}$ GeV.
Suppose the matrix of the
Yukawa couplings of $N_L$, $h_N$, is related to
the matrix of the  Yukawa couplings $h_{\nu}$  of usual doublet $H$ with
$\nu_L$ and $\nu_R$ by relation
\beq
h_{\nu} = S_{L} S_{N}^{-1} h_{N}.
\eeq
(Here $S_{L}$ is the matrix of the transformation which diagonalizes
$h_{\nu}$ and
$S_{N}$ is some almost-arbitrary matrix which may
be related to  features
of $N_{L}$ interactions.
In particular, it can be equal to  $I$ or $S_{L}$).
Then for the light components one gets the following
Majorana mass matrix:
\beq
m^{maj} = \left({v \over \sigma_0}\right)^2 S_N ^T m_N S_N,
\eeq
i.e. at the condition (20)  the smallness  of mixing implied by $m_D$ is
canceled completely
and mixing of light neutrinos is determined by mass matrix
of
superheavy leptons, $m_N$.
One can suggest that
mass terms at  high mass scales have no
hierarchy so that all the elements of matrix  $m_N$ are of
the same order. At $\sigma _0 \sim m_N \sim 10^{16}$ GeV, the
typical mass scale for the lightest components
is about $m \sim v^2 m_N / \sigma_0 ^2 \approx 10^{-2}$ eV.
Small
spread of parameters in $m_N$ allows to explain
 the scales of both the solar  ($ m \sim 0.3 \cdot 10^{-2}$ eV)
and atmospheric ($m \sim (3 - 10) \cdot 10^{-2} {\rm eV}$) neutrino
problems.

The relation between the Yukawa couplings (20)
at $S_L = S_N$ can be reproduced if
$\nu_L$ and $N_L$, as well as $H$ and $\sigma$ enter the
same multiplets.\\

Second possibility implies essentially off-diagonal structure of  $M$ in
the  {\it neutrino Dirac basis}. The letter is determined as a basis
in
which the neutrino Dirac  matrix, $m_D$, is diagonal.
If the  Dirac mass matrices of leptons have the same structure as quark
mass
matrices, the Dirac basis is related to the flavor  basis
by transformation matrix being similar to the  Cabibbo-Kobayashi-Maskawa
matrix.

Special structure of the $M$ in the neutrino Dirac basis can be a
consequence of certain family symmetry G.
 Let us suppose that G = U(1), the usual Higgs doublet is G-neutral,
left and  right-handed neutrino components have the charges
$G(\nu_{i L}) = G(\nu_{i R}) = (1/2,-1/2, 0)$.
Suppose also that model contains  a scalar field $\sigma$ with
charge $G_{\sigma} = 1$ and nonzero vacuum
expectation value $\sigma_0$.
Then the Dirac mass matrix is diagonal
and
 the Majorana mass matrix generated by $\sigma_0$ and  bare mass terms
equals:
\beq
M =
\pmatrix{
h_{11} \sigma_0 &  M_{12} & 0\cr
M_{12} & h_{22} \sigma & 0 \cr
0 & 0 & M_{33} \cr
}\ .
\eeq
The the mixing between 1 and 2 components in Dirac basis
generated by the see-saw with matrix (22) is
$$
\tan 2 \theta_{12}  \approx
2 \frac {m_{1D}}{m_{2D}} \frac {M_{12}}{h_{11} \sigma_0} \ .
$$
If $G$ is a good symmetry at high mass scale, i.e., if $M_{12} \gg
h_{11}\sigma$,
the smallness of mixing related to the
hierarchy of the Dirac mass matrices is compensated. In flavor basis
the  $\nu_e - \nu_{\mu}$ -
mixing will differ from the above by the small angle typical for
Cabibbo-Kobajashi-Maskawa matrix. Changing the prescription of charges
one can get an enhancement of 1 - 3 mixing or both 1 - 2 and 1 - 3  etc. \\

The considered mechanism of  enhancement is realized in SUSY $E_6$
model$^{42}$. All fermions acquire the masses from the
interaction with the same 27-plet of Higgs bosons which means that at tree
level  the Dirac matrices are diagonal and there is no mixing.
In one  version of the model the radiative corrections give only small
nondiagonal elements to the neutrino Dirac mass matrix.  Then it was
postulated that  the Majorana mass matrix, $M_R$, is completely off
diagonal which is dictated by the condition of the calculability
of the theory. Thus, the conditions
considered above are satisfied.\\

Large (maximal) mixing may arise naturally in models with radiative
generation of neutrino masses
, e.g., by Zee mechanism$^{43}$
. \\

It was argued that large mixing can appear if neutrino masses are
generated by ``Planck scale interactions"$^{44, 45}$. Such an interaction
could result in the nonrenormalizable operators of the type
\beq
\frac{\alpha_{ij}}{M_{Pl}} l^T_i l_j H^T H,
\eeq
where $l_i$ are the lepton doublets , and H is the usual Higgs doublet. At
$\alpha \sim 1$ the interaction (23) gives $m \sim H_0^2/M_{Pl} \sim
10^{-5}$ eV and consequently $\Delta m^2 \sim 10^{-10}~ {\rm eV}^2$.
It was postulated in $^{45}$ that in {\it flavor} basis all the couplings
are equal: $\alpha_{ij} = \alpha_0$. Consequently, mass matrix generated by
(24) has the "democratic" form with all the elements being equal to one
value   $m_0$.  Such a matrix
results in two massless states and one massive state with
$m = 3 m_0$. Neutrino $\nu_e$ mixes with the combination
$(\nu_{\mu} - \nu_{\tau})/2$ and the mixing angle defined by
$\sin^2 2 \theta = 8/9$ is precisely in
the region of the "just-so" solution. \\

Large lepton mixing  strongly influences  the neutrino fluxes from
the gravitational collapses of stars. It results in  transitions
 $\bar{\nu}_{\mu} - \bar{\nu}_e$
($\bar{\nu}_{\tau}$).
The crucial point is that the original $\bar{\nu}_{\mu}$
($\bar{\nu}_{\tau}$)
 energy spectrum
has a larger average energy than the spectrum of $\bar{\nu}_e$ (the
opacity of $\bar{\nu}_e$ exceeds that of $\bar{\nu}_{\mu}$).
Then $\bar{\nu}_e \leftrightarrow \bar{\nu}_{\mu}$ transition results in
partial or complete  ``permutation" of the corresponding energy
spectra$^{46}$.
The effect can be described by the permutation factor $p$ (averaged
probability of the conversion). An upper bound
$$
p < 0.35, ~~~ 99\% ~C.L.
$$
has been  derived$^{46}$ using the data from SN1987A and different models of
the
neutrino burst. Considering a propagation of the neutrinos in  matter of
the star, on the way between the star and the Earth, and in  matter of the
Earth, the relation has been found between
the permutation factor and the mixing angle. The upper bound on p
was transformed  in the upper bound on the $\sin^2 2\theta$ (fig. 2).
In particular the excluded region covers the region of  "just-so" solution.

\vskip 0.8cm

\noindent
{\large \bf 5. Conclusions}
\vskip 3mm

1. There are three indications on the existence of nonzero neutrino masses
and mixing.
They are related to the solar,  atmospheric, and relic neutrinos.
Solar neutrino data can be explained by resonant
conversion or/and oscillations with vacuum mixing angles being
or appreciably larger or appreciably smaller than the Cabibbo angle.
Atmospheric neutrino deficit implies the oscillations with large mixing.

2. Solar neutrino data pick up the region of mixing angles
$\theta \sim \sqrt {\frac {m_e}{m{\mu}}}$. This relation
implies the similarity of the lepton and quark mass matrices.
It can follow (as in quark sector) from
the  Fritzsch ansatz.

The above relation of masses and mixing
can be reproduced by the see-saw mechanism of mass generation.
The standard see-saw mechanism for three light neutrinos allows to reconcile
the solution of the solar
neutrino problem via MSW conversion or with explanation of the atmospheric
neutrino deficit or with third neutrino mass being in the
cosmologically interesting region.

3. Large lepton mixing is also quite possible. The enhancement of
lepton mixing due to the see-saw mechanism itself may take place
which implies certain structures of the Majorana
mass matrices.

Maximal and near to maximal  flavor mixing of the electron neutrino is
disfavored   by  the results from SN1987A.

\vskip 0.8cm

\noindent
{\large \bf 6. Acknowledgements}
\vskip 3mm

I would like to thank G.~Senjanovi\'{c}
 for valuable discussions.
\vskip 0.8cm

\noindent
{\large \bf 7. References}
\vskip 3mm

\noindent
1. Los Alamos: R.~G.~H.~Robertson et al., Phys. Rev. Lett. {\bf 67}
(1991) 957. \\
\noindent
2. Mainz: Ch.~Weinheimer et al., Phys. Lett. {\bf B300} (1993) 210.\\
\noindent
3. Livermore: W.~Stoeffl, Bull. Am. Phys. Soc. {\bf 37} (1992) 925.\\
\noindent
4. Zurich: Holtschuh et al., Phys.~Lett. {\bf B287} (1992) 381. \\
\noindent
5. MPI-Moscow: A.~Balysh  et al., Phys. Lett. {\bf B283} (1992) 32;
A.~Pipke, talk given\\
\phantom{hyt} at the International Europhysics Conference on High
Energy Physics, Marseille,\\
\phantom{res} July
22 - 28 (1993). \\
\noindent
6. T.~Bernatowicz et al., Phys.~Rev.~Lett., {\bf 69} (1992) 2341. \\
\noindent
7. M.~Daum et al., Phys. Lett. {\bf B 265} (1991) 425;
M. Janousch et al., Proc. of the\\
\phantom{ujh}Low Energy Muon Science meeting,
April 1993, Santa Fe, New Mexico, USA.\\
\noindent
8. ARGUS: H.~Albrecht et al., Phys. Lett. {\bf B202} (1992) 224.\\
\noindent
9. CLEO: D.~Cinabro et al., Phys. Rev. Lett. {\bf 70} (1993) 3700.\\
\noindent
10. E.~W.~Kolb, M.~S.~Turner, A.~Chakavorty and D.~N.~Schramm, Phys. Rev.
Lett.\\
\phantom{tghy}{\bf 67} (1991) 533.\\
\noindent
11. A.~D.~Dolgov and I.~Z.~Rothstein, Phys.~Rev.~Lett. {\bf 71} (1993) 476.\\
\noindent
12. T.~ Walker, private communication; M.~Kawasaki, G.~Steigman,
H.-S.~Kang,\\
\phantom{iuyt} Nucl. Phys. {\bf B403} (1993) 671. \\
\noindent
13. H.~Pessard,  these Proceedings.  \\
\noindent
14. T.~Ohshima et al., Phys.~Rev. {\bf D47} (1993) 4840.\\
\noindent
15. E.~Holzschuh and W.~Kundig, in Perspectives in Neutrinos, Atomic
Physics, and \\
\phantom{uytr}Gravitation, XIII Moriond Workshop, Villars-sur-Ollon,
Switzerland, January\\
\phantom{tyui}30 - February 6 (1993).\\
\noindent
16. J.~L.~Mortara et al., Phys. Rev. Lett. {\bf 70} (1992) 394.\\
\noindent
17. G.~E.~Berman et al., to submitted to Phys. Rev. Lett. .\\
\noindent
18. M.~Bahran and G.~R.~Kalbfleisch, Phys.~Lett. {\bf B303} (1993) 355;
Phys.~Rev. {\bf D47} \\
\phantom{opi}(1993) R754.\\
\noindent
19. F.~von Feilitzsch, these Proceedings.\\
\noindent
20. S.~Turck-Chi\`{e}ze, these Proceedings.\\
\noindent
21. S.~P.~Mikheyev and A.~Yu.Smirnov, Sov.~J.~Nucl.~
Phys.~{\bf 42 }(1985) 913; Sov.\\
\phantom{agtd}Phys.~JETP {\bf 64} (1986) 4;
 L.~Wolfenstein, Phys.~Rev.~
 D {\bf 17}, 2369 (1978), ibidem,\\
\phantom{yuij}{\bf D20} 2634 (1979).\\
\noindent
22. P.~I.~Krastev and
S.~T.~Petcov, Phys.~Lett. {\bf B299},
 (1993) 99;
 L.~M.~Krauss,\\
\phantom{yujh}E.~Gates and
M.~White, Phys.~Lett. {\bf B298},
 (1993) 94,
Phys.~Rev.~Lett., {\bf 70},\\
\phantom{olki} (1993) 375,
S.~A.~Bludman et al., Phys. Rev. {\bf D47} (1993) 2220; N.~Hata and\\
\phantom{yhgt}P.~Langacker, Univ. of Pennsylvania preprint UPR-0570T;
G.~L.~Fogli,  E.~Lisi\\
\phantom{olki}and D.~Montanino, Preprint CERN-TH 6944/93,
BARI-TH/146-93. \\
\noindent
23. X.~Shi,  D.~N.~Schramm, and
J.~N.~Bahcall, Phys.~Rev.~Lett.~{\bf 69} (1992) 717;\\
\phantom{ert} A.~Yu.~Smirnov, ICTP preprint IC/92/429;
D.~Harley, T.~K.~Kuo and J.~Pan-\\
\phantom{olki}taleone, Phys.~Rev. {\bf D47} (1993)
4059;
A.~S.~Joshipura and P.~I.~Krastev,\\
\phantom{tgvh}Preprint IFP-472-UNC, PRL-TH-93/13.\\
\noindent
24. V.~N.~Gribov and B.~M.~Pontecorvo, Phys.~Lett. {\bf 28}, (1967) 493;
J.~N.~Bahcall \\
\phantom{poil}and  S.~C.~Frautschi, Phys.~Lett. {\bf B29}, (1969) 623;
V.~Barger, R.~J.~N.~Phillips,\\
\phantom{ploi}and K.~Whisnant, Phys.~Rev. {\bf D24},
(1981) 538;
S.~L.~Glashow and L.~M.~Krauss,\\
\phantom{oiuk}Phys.~Lett. {\bf B190},  (1987) 199.\\
\noindent
25.  see for recent analysis  P.~I.~Krastev and S.~T.~Petcov, Phys.~Lett.
{\bf  B285},\\
\phantom{yuih}(1992) 85.\\
\noindent
26. Y.~Totsuka, Nucl.~Phys.~B (Proc. Suppl.) {\bf 31} (1993) 428.\\
\noindent
27. P.~Litchfield, in
International Europhysics Conference on High Energy
Physics,\\
\phantom{loij} Marseille, July
22 - 28 (1993).\\
\noindent
28. E.~L.~Wright et al., Astroph.J. {\bf 396} (1992) L13; P.~Davis,
F.~J.~Summers
and\\
\phantom{vfgb} D.~Schlegel, Nature {\bf 359} (1992) 393;
R.~K.~Schafer and Q.~Shafi, Bartol preprint\\
\phantom{iuyt} BA-92-28, (1992);
J.~A.~Holzman and J.~R.~Primack, Astroph.J. {\bf 405}\\
\phantom{plkj} (1993) 428.\\
\noindent
29. S.~Weinberg, ``A  Festschrift for I.~I.~Raby",
Transaction of the New York Academy\\
\phantom{ikju} of Sci. {\bf 38} (1977) 185; F.~Wilczek and
A.~Zee, Phys.~Lett. {\bf B70} (1977)  418.\\
\noindent
30. H.~Fritzsch, Phys.~Lett. {\bf B70} (1977) 436.\\
\noindent
31. M.~Gell-Mann, P.~Ramond, R.~Slansky, in {\it Supergravity},
ed.~by F.~van Nieuwen-\\
\phantom{poiu}huizen and D.~Freedman (Amsterdam,
North Holland, 1979) 315;\\
\phantom{uyt} T.~Yanagida, in Proc.~of the Workshop on the
{\it Unified Theory and Barion\\
\phantom{ikjh} Number in the Universe}, eds.~
O.~Sawada and A.~Sugamoto (KEK, Tsukuba) 95 \\
\phantom{olki}(1979); R.~N.~Mohapatra and  G.~Senjanovi\'{c},
Phys.~Rev.~Lett.  {\bf 44}, 912 (1980).\\
\noindent
32. H.~Georgi and C.~Jarlskog, Phys.~Lett. {\bf B86} (1979) 297.\\
\noindent
33. E.~Witten, Phys.~Lett. {\bf 91B} (1980) 81.\\
\noindent
34. M.~Fukugita, M.~Tanimoto and T.~Yanagida, Prog.~Theor.~Phys.~
{\bf 89}, 263 (1993).\\
\noindent
35. K.~S.~Babu and Q.~Shafi, Phys. Rev {\bf D47} (1993) 5004. \\
\noindent
36. S.~Dimopoulos, L.~Hall and S.~Raby,
 Phys. Rev. {\bf D45} (1993) R3697.\\
\noindent
37. K.~S.~Babu and Q.~Shafi, Phys.~Lett. {\bf B294} (1992) 235.\\
\noindent
38. K.~S.~Babu and R.~N.~ Mohapatra, Phys. Rev. Lett. {\bf 70} (1993) 2845.\\
\noindent
39. G.~Anderson, S.~Dimopoulos, L.~J.~Hall, S.~Raby and G.~Starkman, LBL
preprint\\
\phantom{tyhg} 33531 (1993).\\
\noindent
40. A.~Yu.~Smirnov, Phys. Rev. {\bf D48} (1993) 3264.\\
\noindent
41. T.~Goldman and G.~T.~Stephenson, Phys. Rev. {\bf D24} (1981) 236;
in Proc. of\\
\phantom{iuyt}Orbis Scientiae (1981), Univ of Miami, Coral Gables,
Florida 12 - 22 January ,\\
\phantom{oiuy} p.111.\\
\noindent
42. Y.~Achiman and A.~Lukas, Nucl. Phys. {\bf 384} (1992) 78;
Phys. Lett. {\bf B296} (1992)\\
\phantom{yujh}128.\\
\noindent
43. A.~Zee, Phys. Lett., {\bf B93} (1980) 389. \\
\noindent
44. R.~Barbieri, J.~Ellis and M.~K.~Gaillard, Phys. Lett. {\bf B90} (1980)
249.\\
\noindent
45. E.~Kh.~Akhmedov, Z.~G.~Berezhiani and G.~Senjanovi\`{c}, Phys.~Rev.~Lett.,
{\bf 69}\\
\phantom{ikln} (1992) 3013.\\
\noindent
46. A.~Yu.~Smirnov, D.~N.~Spergel and J.~N.~Bahcall,
Preprint IASSNS-AST 93/15.

\newpage

\noindent
{\large \bf Figure captions}

\vskip 2mm

Figure 1. Islands on the $\Delta m^2 - \sin^2 2\theta$ plot. Regions of
the neutrino parameters implied by the solar and  atmospheric neutrino
data. Parameters in top island give
the explanation of the atmospheric muon neutrino deficit via
$\nu_{\mu} - \nu_{\tau}$ oscillations.
Dashed lines show the extension of  regions of the MSW-solutions to the
solar neutrino problem, when the effect of third neutrino is taken into
account. Bottom island is the region of the "just-so" solution to the solar
neutrino problem. Dashed-dotted line shows the upper bound on
$\nu_e - \nu_{\tau}$ mixing from SN87A; vertical curves mark
the mixing corresponding to $\theta = \theta_c$ and $\theta = \theta_l$.\\

Figure 2. The textures of the fermion mass matrices. Numbers in
squares are the dimensions of the Higgs boson representations which
generate the corresponding mass terms. F:  Fritzsch ansatz in$^{34}$;
BS: model$^{35}$, DHR: model$^{36}$, BS': model$^{37}$.

\end{document}